# Origin of defects responsible for charge transport in resistive random access memory based on hafnia

D.R. Islamov[1], T.V. Perevalov[1], V.A. Gritsenko[1], V.Sh. Aliev[1], A.A. Saraev[2], V.V. Kaichev[2], E.V. Ivanova[3], M.V. Zamoryanskaya[3], C.H. Cheng[4], A. Chin[5]

[1]A.V. Rzhanov Institute of Semiconductor Physics of SB RAS, 13 Lavrentieva ave., 630090, Novosibirsk, Russia,
[2]Boreskov Institute of Catalysis of SB RAS, 5 Lavrentieva ave., 630090, Novosibirsk, Russia,
[3]Ioffe Physicotechnical Institute, Russian Academy of Sciences, 26 Politechnicheskaya st., 194021, St. Petersburg, Russia,
[4]Dept. of Mechatronic Technology, National Taiwan Normal University, Taipei 106, Taiwan ROC,
[5]Department of Electronics Engineering, National Chiao-Tung University, Hsinchu 300, Taiwan ROC.

## Abstract

A promising candidate for universal memory, which would involve combining the most favourable properties of both high-speed dynamic random access memory (DRAM) and non-volatile flash memory, is resistive random access memory (ReRAM). ReRAM is based on switching back and forth from a high-resistance state (HRS) to a low-resistance state (LRS). ReRAM cells are small, allowing for the creation of memory on the scale of terabits. One of the most promising materials for use as the active medium in resistive memory is hafnia ($HfO_2$).

However, an unresolved physics is the nature of defects and traps that are responsible for the charge transport in HRS state of resistive memory.

In this study, we demonstrated experimentally and theoretically that oxygen vacancies are responsible for the HRS charge transport in resistive memory elements based on $HfO_2$. We also demonstrated that LRS transport occurs through a mechanism described according to percolation theory.

Based on the model of multiphonon tunneling between traps, and assuming that the electron traps are oxygen vacancies, good quantitative agreement between the experimental and theoretical data of current-voltage characteristics were achieved. The thermal excitation energy of the traps in hafnia was determined based on the excitation spectrum and luminescence of the oxygen vacancies.

The findings of this study demonstrate that in resistive memory elements using hafnia, the oxygen vacancies in hafnia play a key role in creating defects in HRS charge transport.

**The major driving force in the current development of microelectronics is the search for a universal memory, which would combine the most favourable properties of both high-speed dynamic random access memory (DRAM) and non-volatile flash memory. The candidates for universal memory are ferroelectric memory, phase-change memory, magneto-resistive memory, and resistive memory. Each option has advantages and disadvantages. One of the most promising options for universal memory is resistive random access memory (ReRAM) [1–5], which has many advantages: a simple metal-insulator-metal (MIM) structure, small memory cell, potential for 3D integration, high read-and-write operation speeds, low power consumption,**

and the ability to store information over the long-term. An ReRAM operation is principally based on switching back and forth from the insulating medium's high resistance state (HRS) to a low resistance state (LRS) when a current flows. Hafnia $HfO_2$ is one of the most promising materials for use as the active medium of resistive memory [6, 7]. Hafnia conductivity is limited by traps ionization [6] when ReRAM is switched to the HRS state. However, the atomic structure of defect that affects the localization and charge transport still remains unclear. Currently, the accepted hypothesis is that oxygen vacancies are responsible for charge transport in dielectric and ReRAM switching [5]. Although many studies have investigated the theory of the atomic and electronic structure of oxygen vacancies in hafnia [8–15], direct experimental data regarding the presence of oxygen vacancies in hafnia have not been reported. Here we clearly show, both experimentally and theoretically, that oxygen vacancies are responsible for HRS charge transport in resistive memory elements based on $HfO_2$. This paper also shows that LRS transport occurs through the mechanism described in percolation theory. This study completely describes the electronic transport in ReRAM insulator medium.

Fig. 1**a** shows a typical hysteresis of experimental current-voltage characteristics of TaN/$HfO_x$/Ni metal-insulator-metal structures at room temperature. The memory window (i.e., current difference), is more than three orders at a voltage bias of 0.5 V. To study the electronic transport mechanism in hafnia, we performed transport measurements in simpler metal-insulator-semiconductor (MIS) structures, specifically $n$-Si/$HfO_x$/Ni. The experimental current-voltage characteristics in $n$-Si/$HfO_x$/Ni structures, measured at different temperatures with a positive bias applied to the Ni contact, are shown in Fig. 1**b**, graphed by characters in different colors. We attempted to qualitatively explain the experimental data by using different models, such as the Frenkel trap ionization model [16], the Hill model of trap ionisation [17], and the model of multiphonon trap ionization [18]. However, the fitting procedures involved in using these models returned the nonphysical fitting parameter values.

To describe the experiments quantitatively and qualitatively, we performed simulations based on the model of phonon-assisted tunnelling between traps [19]:

$$J = eN_t^{\frac{2}{3}}P,$$
$$P = 2\frac{\sqrt{\pi}\hbar W_t}{m^*s^2\sqrt{2(W_{opt}-W_t)}}\exp\left(-\frac{W_{opt}-W_t}{2kT}\right)\exp\left(-\frac{2s\sqrt{m^*W_t}}{\hbar}\right)\sinh\left(\frac{eFs}{2kT}\right),$$ (1)

where $J$ is the current density, $e$ is the electron charge, $P$ is the probability of electron tunnelling between traps per second, $\hbar$ is the Planck constant, $W_t$ is the energy of thermal excitation of the trap, $W_{opt}$ is the energy of optical excitation of the trap, $s = N_t^{-1/3}$ is the distance between traps, $k$ is the Boltzmann constant, and $F$ is the electric field. The results of this multi-parameter fitting procedure are shown in Fig. 1**b**, graphed in solid lines. This procedure yielded the values of different transport parameters, such as $N_t = 2.5 \times 10^{20}$ cm$^{-3}$, $W_t = 1.25$ eV, $W_{opt} = 2W_t = 2.5$ eV, and $m^* = 0.8\, m_e$. Quantitatively, there is full agreement between the model of phonon-assisted tunnelling between traps and the experimental data. The electron trap energy in hafnium oxide was previously determined in [20] to be 1.2 eV. The trap thermal energy value of 1.25 eV that we obtained is close to that of $W_t = 1.36$ eV observed in the experiment [21].

Furthermore, the trap optical energy value of $W_{opt}$ = 2.5 eV is close to the calculated value of 2.35 eV for the negatively charged oxygen vacancy in hafnia reported in [14].

Fig. 2**a** shows the configuration diagram of a negatively charged oxygen vacancy (electron trap) in hafnia. A vertical transition with a value of 2.5 eV corresponds to the optical trap excitation. The thermal trap energy value was 1.25 eV.

We applied the same fitting procedures to the data obtained in our experiments of current-voltage characteristics in MIM structures. Fig. 3 shows the experimental data (graphed in colour) of current-voltage measurements at different temperatures. The solid-coloured lines present the results of HRS simulations of phonon-assisted tunnelling between traps (Eq. (1)). MIM structures have the following transport parameter values: $N_t$ = 5.5×10$^{20}$ cm$^{-3}$, $W_t$ = 1.25 eV, $W_{opt}$ = 2.5 eV = 2$W_t$, $m^*$ = 0.9 $m_e$. The difference between MIM and MIS structures in effective mass is due to bulk charge, which is adequately addressed in [22]. However, it is important to notice that the trap's energy parameters are invariants of grown structures. Consequently, we found that the nature of charge carrier transport in the HRS of ReRAM structures is phonon-assisted tunnelling between traps.

To examine the defects in atomic and electron structures, we measured the X-ray photoelectron spectra (XPS) of the valence band, cathodoluminescence (CL), and electron energy loss spectra (EELS) in nonstoichiometric hafnia. The experimental XPS spectra of the valence band obtained before and after Ar$^+$ ion bombarding the stoichiometric amorphous Si/HfO$_2$ film, as well as the spectra of nonstoichiometric Si/HfO$_{x<2}$, are shown in Fig. 4**a**–**c**. The valence band XPS spectra for both HfO$_x$ films have a rather intense peak, nearly 3 eV above the top of the hafnia valence band. The peaks in this region might have been caused by electron photoemission from the 5d level of metallic hafnium. The Ar$^+$ ions irradiating Si/HfO$_x$ have an additional XPS peak, 1 eV higher than the top of the valence band. According to our calculations, the peak is related neither to neutral oxygen polyvacancies nor to charged monovacancy. The calculated total density of states (TDOS) of the valence band for c-HfO$_2$, in considering oxygen monovacancy, agrees with the valence band width in a defect level position (a feature of TDOS at 3.0 eV in Fig. 4**d**). These results prove the presence of oxygen vacancies in amorphous nonstoichiometric hafnia and hafnia films following Ar$^+$ ion bombardment.

Experimental CL and EELS in nonstoichiometric HfO$_{1.9}$ are shown in Fig. 4**e**–**h**. A peak with a value of 2.6 eV was observed in the CL HfO$_x$ spectrum (Fig. 4**e**). Previously, a cathodoluminescence band with a value of 2.7 eV was observed in [23]. Excitation in the EELS (Fig. 4**f**) is observed in HfO$_{1.9}$ at 5.0 eV, although the margin of error for determining EELS peak energy values is 0.3 eV. The photoluminescence (PL) spectrum and excitation photoluminescence band spectrum 2.7 eV (PLE) in HfO$_x$, obtained from the scientific literature [24], are also presented in Fig. 4**g** and Fig. 4**h**, respectively. A peak of 2.7 eV is also observed in the PL spectrum of HfO$_x$. Luminescence band excitation with a maximum value of 2.7 eV is located at 5.2 eV.

To identify the defect responsible for excitation at 5.2 eV, we examined the optical absorption of an oxygen vacancy in hafnium oxide. Fig. 4**i** shows a calculated hafnium oxide optical absorption spectrum. The dominant peak has a value of 5.3 eV. The oxygen vacancy excitation corresponds to an optical transition of 5.3 eV. The calculation shows a wide absorption spectrum with energy values between 4.7–4.8 eV. Optical absorption in hafnium oxide with similar energy values was experimentally observed in [20]. A previous quantum mechanical simulation established that absorption in hafnium oxide within 4.7–4.8 eV is connected to the transition of electrons from the top of the valence band to the excited state of oxygen vacancy [10].

A diagram of the configuration of neutral oxygen vacancy is shown in Fig. 2**b**. A vertical transition with an energy value of 5.2 eV corresponds to optical electron excitation, and a transition with an energy value of 2.7 eV corresponds to the radiating transition into equilibrium. The energy distributed to phonons corresponds, in our model, to the thermal trap energy.

These data indicate that the luminescence band of 2.7 eV and the excitation band of 5.2 eV are conditioned by own defects, namely, oxygen vacancies in $HfO_{1.9}$.

The most commonly used LRS model in ReRAM structures consists of a metal filament of approximately 10 nm [25, 26]. This metal filament must have linear current-voltage characteristics. This approach exhibits two bottlenecks: (a) it cannot describe ReRAM in $GeO_x$ and $SiO_x$-based structures [27, 28], and (b) the experimentally measured current is much lower than that predicted by Ohm's law. To illustrate the second case, we calculated the current through a pure Hf wire, 10 nm in diameter and 8 nm long. The results are shown in Fig. 3 by a dashed green line. The calculations are approximately $10^5$ times higher than those obtained in the experiments.

The LRS conductivity of theoretically calculated metal filament (pure hafnium, 10 nm in diameter) is 5 orders higher than the measured conductivity. Therefore, we suppose that LRS conductivity is conditioned by the presence of the nonstoichiometric $HfO_x$ phase. The outlook on the structure of nonstoichiometric oxides and nitrides was developed in [29-32] for $SiO_x$, $SiN_x$ and $SiO_xN_y$. In the random mixture (RM) model, $SiO_x$ consists of a mixture of two phases, Si and $SiO_2$. In the random bonding (RB) model, $SiO_x$ consists of a mixture of five types of tetrahedron (Si-$Si_4$, Si-$SiO_3$, Si-$Si_2O_2$, Si-$Si_3O$, and Si-$O_4$), the statistics of which are described using binominal distribution. An intermediate $SiO_x$ structural model is proposed in [30]. According to the intermediate model (IM), $SiO_x$ consists, as in the RB model, of five types of tetrahedron; however, tetrahedron distribution is not described according to random binominal distribution.

A 2D structural image of nonstoichiometric $HfO_x$, regarding the IM, is presented in Fig. 5**a**. According to this model, the filament in hafnium oxide consists of a mixture of metallic hafnium clusters sprinkled into nonstoichiometric $HfO_x$. Fig. 5**c** is an energy diagram of $HfO_x$ in the IM model. According to this diagram, spatial fluctuations of $HfO_x$ chemical composition lead to local band gap width spatial fluctuations. The maximal fluctuation of the energy scale is equal to the $HfO_2$ band gap width of $E_g$=5.6 eV [33]. The work function of metallic hafnium is 4.0 eV. The maximal fluctuation scale of the $HfO_x$ conduction band is 2.0 eV that equals the electron barrier height of Hf/$HfO_2$. The maximal fluctuation scale of the $HfO_x$ valence band is 2.5 eV, the hole barrier height of Hf/$HfO_2$ (Fig. 5**b**).

The large-scale fluctuations of the bottom of conduction band $E_c$ and the top of valence band $E_v$ are close to the model developed by A. L. Éfros and B. I. Shklovskii [34, 35], as shown in Fig. 5**d**. The charge transport in such electron systems can be described according to percolation. This model assumes that excited electrons are thrown to flow level $E^e$ and, rounding a random potential, they transfer the charge. The hole conductivity is realized through the excitation of electrons with energy $E^h$ to the Fermi level with the formation of a hole-type quasi-particle with a further charge transfer according to the obtained quasi-particle. In other words, to be involved in transport processes, electrons and holes must overcome energy thresholds ($W^{e,h}$ here and $W^e \neq W^h$ in general).

The current-voltage characteristics are exponentials [34]:

$$I = I_0(T) \exp\left(\frac{(CeFaV_0^v)^{\frac{1}{1+v}}}{kT}\right), \qquad (2.1)$$

$$I_0(T) \sim \exp\left(-\frac{W}{kT}\right), \qquad (2.2)$$

$$kT\left(\frac{kT}{V_0}\right)^v \ll eFa \ll V_0, \\ kT \ll V_0, \qquad (2.3)$$

where $a$ is the space scale of fluctuations, $V_0$ is the amplitude of energy fluctuation, $C$ is a numeric constant, and $v$ is a critical index. The values of the constants were derived from Monte-Carlo simulations and evaluated at C ~ 0.25 [35], $v$ = 0.9 [34].

Fig. 6 illustrates the fitting procedure for LRS simulations from using the percolation model. The current-voltage characteristics of ln($I$) vs. $F^{1/(1+v)}$ plate (characters), shown in Fig. 6a, have two linear parts: a low-field part (red line) and a high-field part (blue line). We found that the low-field linear part corresponds to the model when limits of equations (2.3) are applied. The results of LRS simulations regarding the percolation model are shown in Fig. 4 by dotted coloured lines. Numeric fitting returns the value of combination as $CaV_0^v$ = 0.45 nm eV$^v$, which corresponds to $V_0$ = 1.9 eV when $a$ = 1 nm and $C$ = 0.25. Fig. 6b shows the simulations in pre-exponent $I_0(T)$ from equation (2.1) (represented by characters). The slope of a fitting line, in red, corresponds to a percolation threshold of $W$ = 460 meV. Assuming that $V_0 \sim W$ enables evaluating the space scale of fluctuations as $a$ ~ 3–4 nm.

The energy threshold for the charge carriers can be evaluated using the calculated current-voltage characteristics of a filament and experimental LRS data. Assuming $I_{LRS}= I_{filament}\times\exp(-W^*/kT)$, this yields an effective activation energy threshold $W^*$ of ~300 meV. Although this was unexpected, the energy threshold values obtained using different approaches yielded close results.

We also attempted to simulate the LRS regarding phonon-assisted tunnelling between traps, as we did for an HRS. The results are shown in Fig. 1, by a solid green line. Using the same values for energy and effective mass obtained from HRS simulations, we found that the results from the LRS simulations and the experimental data were in a very good agreement by using a value of $N_{LRS}$ = 1.1×10$^{22}$ cm$^{-3}$. We understand that the trap concentration is close to the density of atoms in metals ~10$^{22}$ cm$^{-3}$, and that considering vacancies with such a high concentration is inappropriate; hence, the obtained $N_{LRS}$ value is outside the model limits. However, surprisingly, phonon-assisted tunnelling between traps formally describes LRS conduction.

Previous experiments in charge transfer have demonstrated that hafnium oxide conductivity is bipolar (or two-band) [36–38]; electrons are injected from a negatively shifted contact in the dielectric, and holes are injected from a positively shifted electrode in the dielectric. HfO$_2$ conductivity in an HRS was examined in these prior experiments. In our model, filament conductivity in LRS can also be studied using electrons and holes. For reasons of simplicity, the current study was limited to considering monopolar electron conductivity both in the HRS and LRS.

To summarize, we examined the transport mechanisms of hafnia, demonstrating that transport in an HRS is described according to phonon-assisted tunnelling between traps. Simulating the current-voltage characteristics of this model and comparing experimental data with calculations revealed the energy parameters of the traps in hafnia: a thermal excitation energy of 1.25 eV and an optical excitation energy of 2.5 eV.

Oxygen vacancies and a vacancy-conditioned luminescence band of 2.7 eV were identified in the current study by using hafnium oxide photoelectron spectroscopy and quantum-chemical simulations. Photoelectron valence band spectra in stoichiometric $HfO_2$ and slightly nonstoichiometric $HfO_{x<2}$ were investigated. The photo-electron spectroscopy data revealed electron states in the $HfO_x$ band gap. Quantum mechanical simulation indicated that these states are conditioned by oxygen vacancies. These simulations established that the cathodoluminescence band of 2.7 eV in $HfO_x$ and the excitation peak of 5.2 eV are also conditioned by oxygen vacancies.

These results facilitated determining that oxygen vacancies act as charge carrier traps. We also demonstrated that charge transport in LRS is described according to the percolation model in electron systems exhibiting potential large-scale fluctuations.

Our results will be used in producing a higher-quality ReRAM device in the future.

**Methods**

Our devices were fabricated as follows. $HfO_{x<2}$ films were produced using ion beam sputtering deposition. For substrates, we used silicon plates, Si (100), with a resistivity of 4.5 Ω×cm, deep cleaned using a technique by the RCA Company. The target was sputtered by a beam of $Ar^+$ ions, while highly pure oxygen ($O_2$ > 99.999%) was simultaneously delivered into the area near the target and substrate. The composition (x-parameter) of the $HfO_x$ film was defined using the gas flow controller to control the partial oxygen pressure. In our experiments, we grew two sets of $HfO_x$ samples at partial oxygen pressures $9\times10^{-3}$ Pa and $2\times10^{-3}$ Pa. Under such conditions, we produced samples of near-stoichiometric (x ≈ 2) and nonstoichiometric compositions (x < 2).

The valence band spectra of grown hafnia films were obtained using a SPECS XPS machine with a monochromatic Al Kα radiation ($h\nu$ = 1486.74 eV). Defects in $HfO_2$/Si films were generated by $Ar^+$ ion bombardment with an ion energy of 2.4 keV and an ion current density of approximately 10 μA/cm$^2$. It is known that the sputtering by $Ar^+$ preferentially removes oxygen from hafnium oxides leading to the formation of hafnia suboxides ($HfO_{x<2}$) [39]. The [O]/[Hf] atomic ratio was estimated according to the XPS peak areas combined with the Scofield cross-sections. The EELS were measured as described in [40]. The [O]/[Hf] atomic ratio for nonstoichiometric and $Ar^+$-bombarded hafnia film is approximately 1.9.

The CL spectra of the $HfO_x$ samples were conducted using the Camebax electron probe microanalyser. The CL spectra were recorded in the 1.5 eV to 3.8 eV range at 300 K. The sample was irradiated by a beam of 1000 eV electrons, penetrating to a depth of 0.2 μm with a current density of 1.2 A/cm$^2$.

Transport measurements were performed in MIS and MIM structures. For the MIS Si/HfO$_x$/Ni structures, the 20-nm-thick amorphous hafnia (HfO$_{x<2}$) was deposited on an *n*-type Si wafer by using the atomic layer deposition (ALD) system. Tetrakis dimethyl amino hafnium (TDMAHf) and water vapor were used as precursors at a chamber temperature of 250°C for HfO$_x$ film deposition. For the ReRAM measurements, the MIM structures of Si/TaN/HfO$_x$/Ni were used. To fabricate these structures, we deposited the 8-nm-thick amorphous hafnia on 100-nm-thick TaN films, and the films on Si wafers, using physical vapour deposition (PVD). A pure HfO$_2$ target was bombarded by an electron beam, and HfO$_2$ molecules were deposited on the wafer. We did not apply any post-deposition annealing (PDA) to produce the most nonstoichiometric films. Structural analysis showed that the resulting HfO$_x$ films were amorphous. All samples for transport measurements were equipped with round 50-nm-thick Ni gates with a radius of 70 μm.

Transport measurements were performed using a Hewlett Packard 4155B semiconductor parameter analyzer and an Agilent E4980A precision LCR meter. The semiconductor parameter analyser was protected against short circuiting (which would limit the current through the sample) by 1 μA.

The *ab initio* simulations of electronic and optical properties were performed based on the density functional theory (DFT) by using Quantum-ESPRESSO software [41]. We adopted a hybrid functional B3LYP to reproduce the accurate band gap values. We used 81-atom cubic hafnia (c-HfO$_2$) supercells for simulating the oxygen vacancy. The choice of the cubic phase is justified according to its simple structure.

**Supplementary Information** is available in the online version of the paper.

**Acknowledgements.** This work was supported by projects #18.24 of the Siberian Branch of the Russian Academy of Sciences, by the project #1.13 of the Siberian Branch of the Russian Academy of Sciences and the National Science Council, Taiwan, under grant #NSC-100-2923-E-009-001-MY3, by the project #12-08-31084-mol_a of the Russian Foundation for Basic Research. The computations have been done at the Novosibirsk State University Supercomputer Center. The authors would like to thank Andrey E. Dolbak for his assistance with running the EELS experiments.

**Author Contributions** V.A.G. and A.C. planned the project and supervised the research. C.H.C., A.C. and V.Sh.A grew the materials. C.H.C. and A.C. fabricated the MIM and MIS devices and collected the data of transport measurements. A.A.S. and V.V.K. collected the data of photoelectron spectra. E.V.I. and M.V.Z. collected the data of cathodoluminescence spectra. T.V.P computed electronic and optical spectra. D.R.I., V.A.G., T.V.P., and A.C. analyzed and discussed the data. T.V.P, V.A.G. and D.R.I. wrote the manuscript, which was edited and approved by all co-authors.

**Author Information** Reprints and permissions information is available at www.nature.com/reprints. The authors declare no competing financial interests: details are available in the online version of the paper. Readers are welcome to comment on the online version of the paper. Correspondence and requests for materials should be addressed to I.D.R. (damir@isp.nsc.ru), V.A.G. (grits@isp.nsc.ru) and A.C. (albert_achin@hotmail.com).

**Figure Captures**

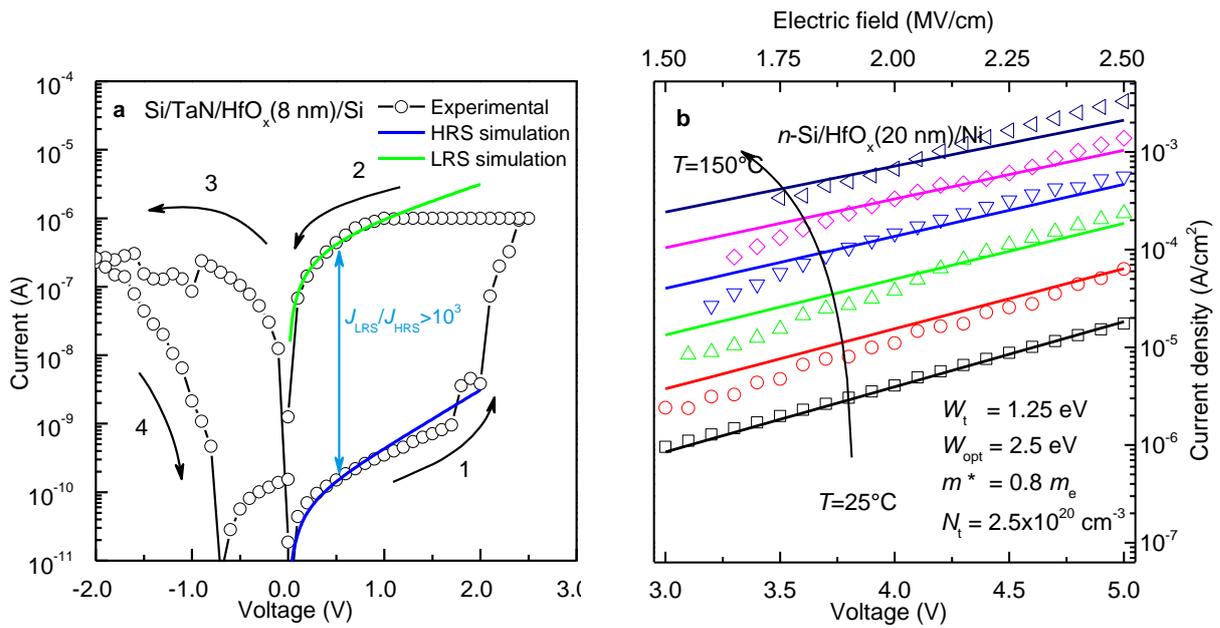

Fig. 1. Experimental current-voltage characteristics (characters) and simulations (lines). **a** ReRAM current-voltage characteristics hysteresis in Si/TaN/HfO$_x$/Ni MIM structures at room temperature. Blue solid line represents HRS simulation by phonon-assisted tunneling between traps at room temperature. Green solid line shows LRS simulation by phonon-assisted tunneling between traps at room temperature. Simulation procedure gives the following parameters of HfO$_x$: $W_{opt}$ = 2.5 eV, $W_t$ = 1.25 eV, $N_{HRS}$ = 1.9×10$^{21}$ cm$^{-3}$, $N_{LRS}$ = 1.1×10$^{22}$ cm$^{-3}$. **b** Experimental Current-Voltage characteristics of $n$-Si/HfO$_x$/Ni structure and simulation by phonon-assisted tunneling between traps (Eq. (1)) at different temperatures $T$ = 25–150°C. Δ$T$=25°C.

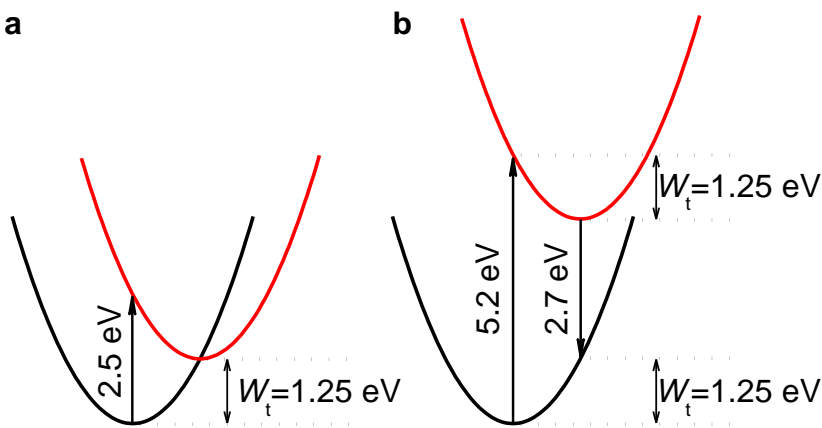

Fig. 2. Configuration coordination energy diagrams of the trap. **a** The diagram of trap ionization process on negative charged oxygen vacancy in hafnia (black line – ground filled state, red line – excited empty state). **b** The diagram of optical transition on neutral oxygen vacancy in hafnia (black line - ground filled state, red line – excited filled state).

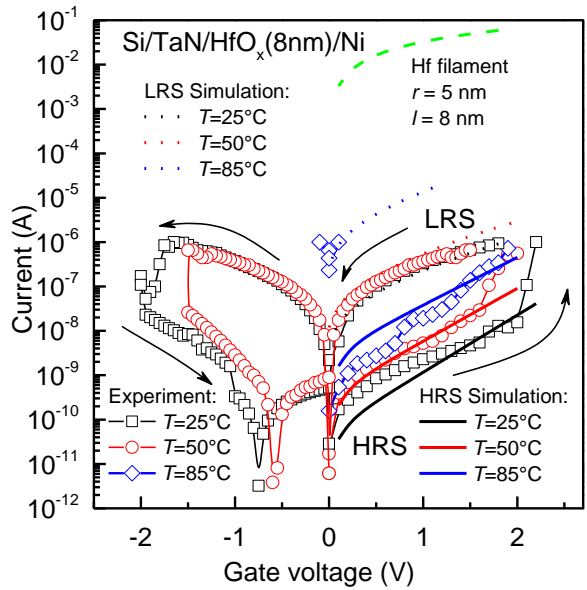

**Fig. 3. Experimental ReRAM current-voltage characteristics hysteresis (characters) in Si/TaN/HfO$_x$/Ni structures at different temperatures.** Black, red and blue solid lines represent HRS simulation by phonon-assisted tunneling between traps at room temperature with the following parameter values: $W_{opt}$ = 2.5 eV, $W_t$ = 1.25 eV, $N_t$ = 5.5×10$^{20}$ cm$^{-3}$, $m^*$ = 0.9 $m_e$. Black, red and blue dotted lines resent LRS simulations in terms of percolation model. Green dashed line models current-voltage characteristics of pure Hf filament of diameter of 10 nm and length of 8 nm.

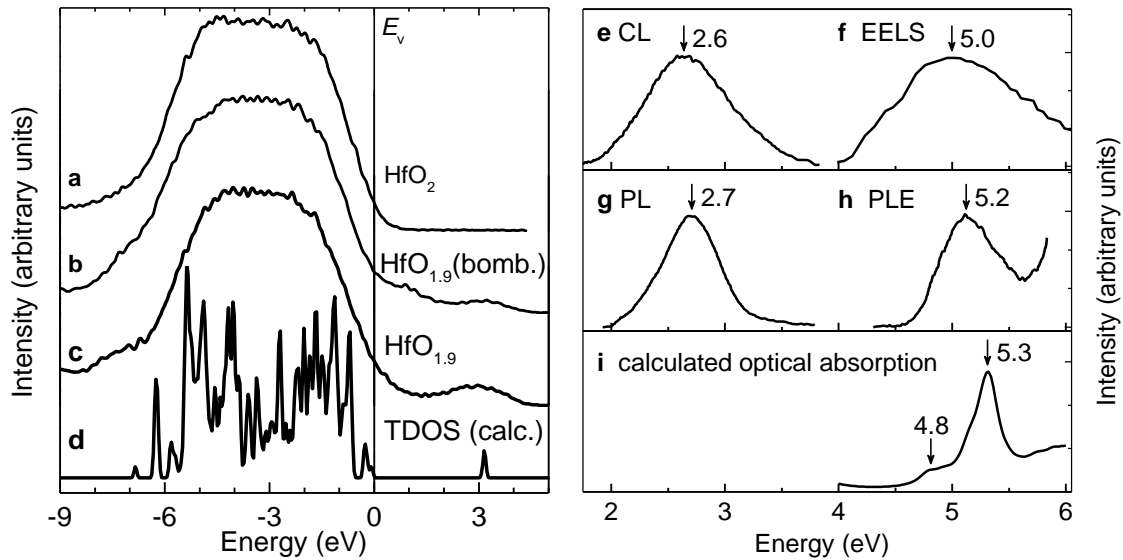

**Fig. 4. Experimental and calculated spectra in hafnia. a** Experimental valence band X-ray photoelectron spectra for amorphous stoichiometric HfO$_2$, **b** bombed hafnia (HfO$_{1.9}$) and **c** nonstoichiometric hafnia (HfO$_{1.9}$). **d** The calculated TDOS for c-HfO$_2$ with oxygen vacancy. **e** Cathodoluminescence spectra of HfO$_{1.9}$. **f** Electron energy loss spectra of HfO$_{1.9}$. **g** Photoluminescence spectra. **h** Photoluminescence excitation spectra of HfO$_{x<2}$ from [24]. **i** Calculated optical absorption spectra for hafnia with oxygen vacancy.

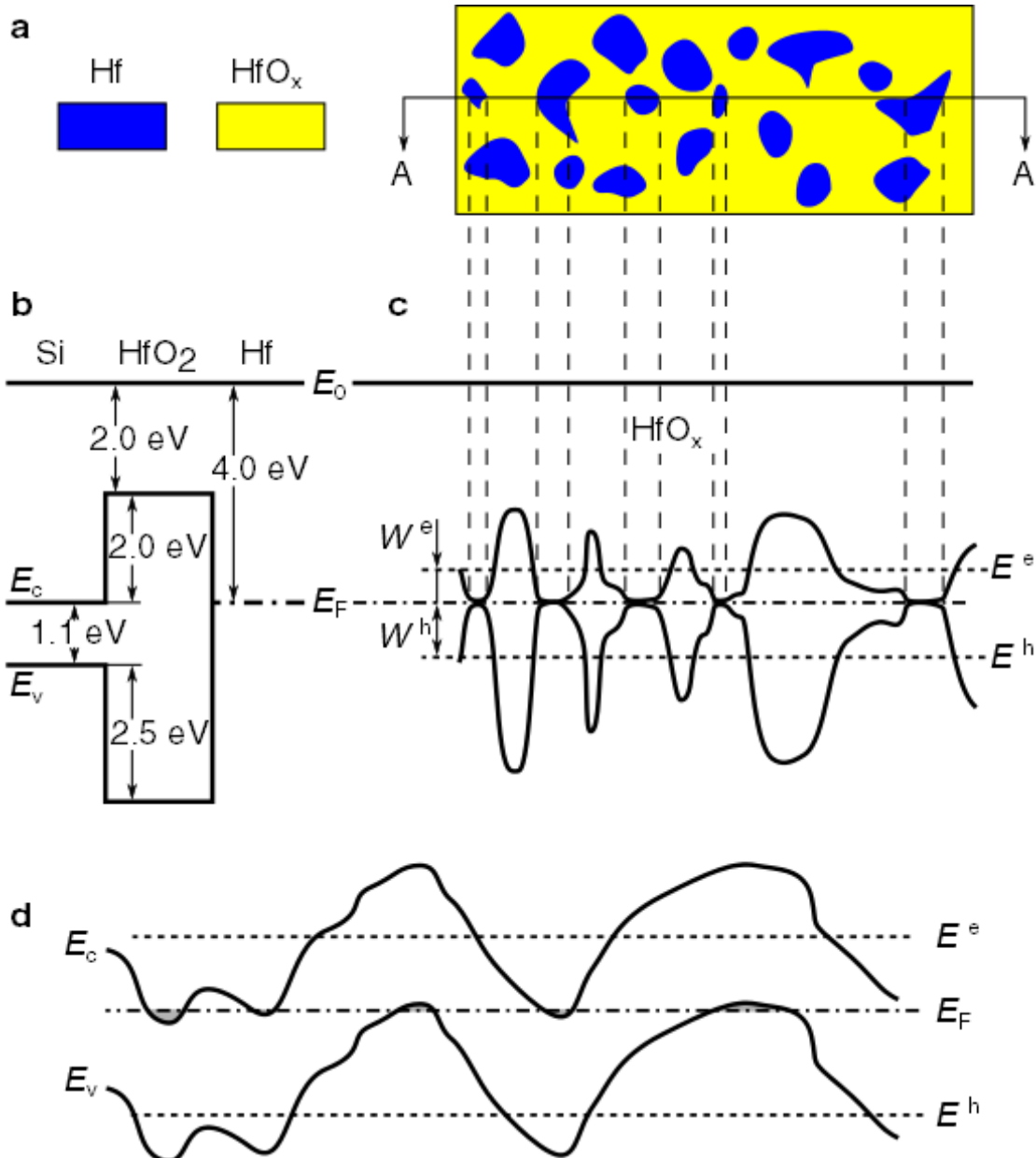

**Fig. 5. Percolation model in electron systems with the large-scale potential fluctuations.** **a** Schematic planar illustration of Hf/HfO$_x$ (x ≪ 2) space-modulated by chemical composition structure. **b** Flat band energy diagram of Si/HfO$_2$/Hf structure. **c** Energy diagram of the structure with large-scale potential fluctuations. **d** Energy diagram of large-scale potential fluctuations in a semiconductor layer [35]. $E^{e,h}$ and $W^{e,h}$ are percolation levels and percolation threshold of the electrons and holes respectively.

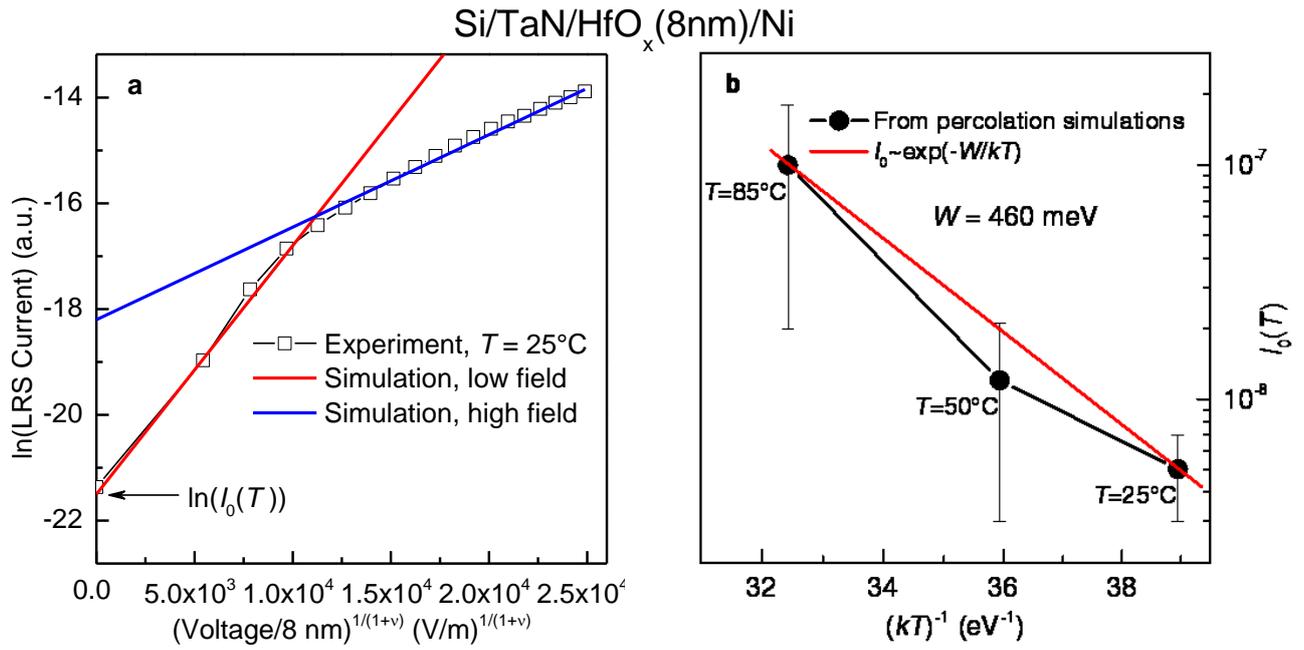

**Fig. 6. Illustration of LRS simulations in terms of percolation model. a** The characters show experimental data, red solid line represents simulations of low-field part of current-voltage characteristics, blue solid line represents high-field part. Critical index $v$ = 0.9. **b** The characters show $I_0(T)$ found from fitting of Eq. (2.1), red solid line represents simulations by equation (2.2) using value of percolation threshold $W$ = 460 meV.